\documentclass{ECOC-author}
\usepackage[position=below]{subfig}
\usepackage{hyperref}
\hypersetup
{
    pdftitle={ECOC-2019:FSO-PCS},
    pdfauthor={Fernando Guiomar},
    colorlinks=false,
    linkcolor=black,
    citecolor=black,
    hidelinks,
    pdfview=Fit,
    pdfpagemode=UseNone
}

\begin{document}

    \title{HIGH-CAPACITY AND RAIN-RESILIENT FREE-SPACE OPTICS LINK ENABLED BY TIME-ADAPTIVE PROBABILISTIC SHAPING}

\author{F. P. Guiomar\ad{1}\corr, A. Lorences-Riesgo\ad{1}, D. Ranzal\ad{2}, F. Rocco\ad{1,3}, \\
A. N. Sousa\ad{1}, A. Carena\ad{3}, A. L. Teixeira\ad{1,2}, M. C. R. Medeiros\ad{1}, P. P. Monteiro\ad{1,2}}

\address{\add{1}{Instituto de
Telecomunica\c{c}\~{o}es, 3810-193, Aveiro, Portugal}
\add{2}{Dep. of Electronics, Telecommunications, and Informatics (DETI), University of Aveiro, 3810-193, Portugal}
\add{3}{DET, Politecnico di Torino, Corso Duca degli Abruzzi, 24, 10129, Torino, Italy}
\email{guiomar@av.it.pt}}
    

\begin{abstract}
\vspace{-0.3cm}
Using time-adaptive probabilistic shaped 64QAM driven by a simple SNR prediction algorithm, we demonstrate $>$450 Gbps transmission over a 55-m free-space optics link with enhanced resilience towards rainy weather conditions. Through continuous measurement over 3-hours, we demonstrate $\sim$40~Gbps average bit-rate gain over unsupervised fixed modulation.
\end{abstract}
    \maketitle
    
    \section{Introduction}

The use of free-space optics (FSO) for high-speed communications has recently been attracting a strong interest among academic and industrial players, currently finding wide applicability for indoor \cite{FSO-MIMO-2018} and inter-satellite communications \cite{FSO-satellite-2017}. 
In addition, FSO can also be considered as a wireless alternative to fiber-based transmission in terrestrial applications \cite{FSO-realTime-2018,FSO-KK-2019}, being a promising candidate for ultra-high capacity fronthauling in beyond 5G access networks \cite{FSO-vertFH-2018}. 
However, the robustness of FSO against adverse weather conditions on outdoor terrestrial deployments still lacks further experimental validation \cite{FSO-weather-2015,FSO-turbulence-2016}. 
On the contrary of mm-wave frequencies (30--300~GHz), which suffer from serious rain fading issues \cite{mmW-rain-2015}, FSO is potentially more robust to rain-induced attenuation, since the radius of raindrops ($>100$~$\mu$m) is much larger than the transmitted optical wavelengths. 
Nevertheless, some attenuation of the signal is still expectable due to the non-selective scattering of light when passing through raindrops \cite{FSO-weather-2015}. 
For most applications, this additional attenuation must be precisely taken into account in order to maximize the channel capacity. Therefore, it urges to develop flexible and robust communication techniques that might self-adapt to the evolving FSO channel conditions in outdoor deployments. 

Probabilistic constellation shaping (PCS) has recently been proposed for optical communications systems \cite{PCS-2016-Buchali}, and has ever since seen a rapid dissemination of both research and commercial applications \cite{PCS-2019-Bocherer,PCS-2019-Cho}. 
The key advantages of PCS that leveraged its recent success in optical communications are mainly twofold: i) the so-called shaping gain over uniform QAM signaling, which can theoretically amount up to 1.53~dB \cite{PCS-shapingGain-1993}, and ii) the enhanced flexibility that it provides for the adaptation of bit-rate with arbitrary granularity \cite{PCS-flexibility-2017}. 
Taking advantage of this bit-rate flexibility feature of PCS, in this work we experimentally demonstrate a time-adaptive modulation scheme to continuously adjust the transmission rate and maximize the channel throughput of a 55-m outdoor FSO link.

    \section{Experimental Setup}

\begin{figure*}[t!]
    \centering
    \includegraphics[width=16cm]{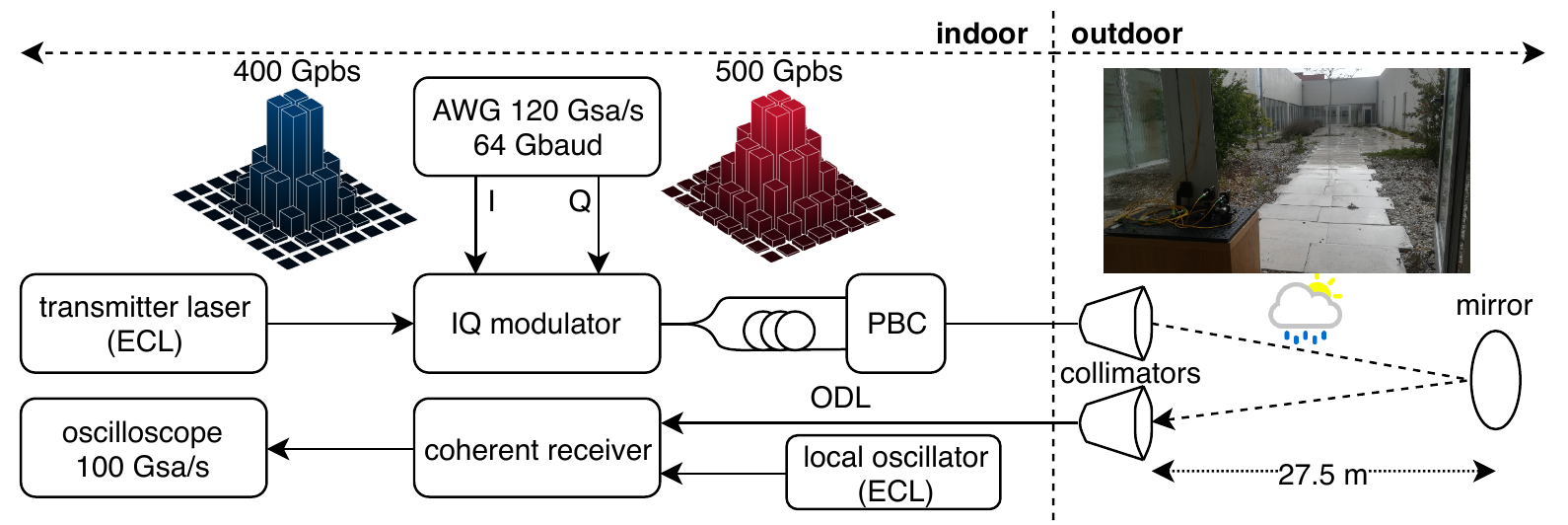}
    \caption{Experimental setup for optical signal transmission and detection with a 55-m free-space optics link.}
    \label{fig:expSetup}
\end{figure*}

The experimental setup utilized in this work is depicted in \figurename.~\ref{fig:expSetup}.
The baseband electrical signal is digitally modulated at 64~Gbaud and generated by an arbitrary waveform generator (AWG) operating at 120~Gsa/s, with 45~GHz of analog bandwidth.
In order to enable a flexible bit-rate, we use PCS with a 64QAM template and the constant composition distribution matcher \cite{PCS-CCDM-2016} to adapt the symbol probability to the desired transmitted bit-rate. 
In the transmitted signal, we allocate 20\% overhead (rate 5/6) for forward error correction (FEC) and 6.25\% overhead (rate 15/16) for DSP pilots, thus yielding a \textit{net symbol-rate} of 50~GBaud, which corresponds to a maximum of 600~Gbps net bit-rate when the dual-pol PCS-64QAM constellation is loaded with its maximum entropy of 6~bits/symbol (which leads to standard uniform 64QAM signaling). 
The DSP pilots are QPSK-like symbols inserted at the average signal power of the PCS constellation, in order to avoid modifying the overall transmitted power. 
This signal is then optically modulated in a single-polarization IQ modulator with $\sim$22~GHz bandwidth. The optical source is an external cavity laser emitting at 1550~nm with $\sim$100~kHz linewidth and 15~dBm optical power.
The dual-polarization optical signal is generated using an optical delay line with 1~m, thus ensuring time decorrelation between the polarization tributaries. 
The optical signal is then sent to the FSO link, which is composed of two collimators with $0.017^\circ$ full-angle divergence and a 24~mm  diameter.
As depicted in \figurename.~\ref{fig:expSetup}, a concave mirror with $\sim$30~m focal distance is utilized to double the transmission length, leading to a total of 55~m.
It is important to mention that both the collimators and the mirror are fully exposed to the outdoor weather without any special protection. 
At the receiver, the optical signal is converted to the electrical domain in a dual-polarization coherent receiver with $\sim$40~GHz bandwidth, where the signal is mixed with a local oscillator with $\sim$100~kHz linewidth and 13~dBm optical power. 
Note that, since we consider the pratical case of unamplified optical transmission, the performance of this system is primarily dictated by the received optical power, which will strongly depend on the time-varying FSO link attenuation. 
Each I and Q component is then sampled and digitized by a 4-port real-time oscilloscope with 100~Gsa/s and 33~GHz analog bandwidth.
Finally, the captured waveforms are collected in MATLAB where offline digital signal processing (DSP) is applied to compensate for the channel impairments. 

The DSP subsystem is initiated with the compensation of the receiver frontend imbalance using Gram-Schmidt orthogonalization \cite{DSP-GS-2006}. 
Polarization demultiplexing and a first coarse channel equalization is performed by the constant modulus algorithm (CMA) with 25 taps in a $2\times 2$ butterfly configuration. An initial fully data-aided training stage is applied for convergence of the equalizer, which is then switched to pilot-based equalization. 
After frequency recovery, carrier-phase estimation is performed by estimating the phase on the QPSK pilot symbols and interpolating to the remaining samples \cite{DSP-pilotCPE-2012}. 
A second adaptive least-mean squares (LMS) equalizer with 51~taps and $4\times 4$ real-valued filters for the in-phase and quadrature components in both polarizations is then applied to fine tune the equalization of the signal, also compensating for the residual IQ skew \cite{DSP-LMS4x4-2013}. 
Finally, the signal is downsampled and demapped, and the normalized generalized mutual information (NGMI) is calculated utilizing the procedure described in \cite{DSP-NGMI-2017}.
In addition, we also evaluate the error vector magnitude (EVM) between the fully-processed received signal and the transmitted one, from which we can infer the received signal-to-noise ratio (SNR), by simply applying the relation $\mathrm{SNR_\mathrm{dB}} = -20\log_{10}(\mathrm{EVM_\%}/100)$.

    \section{Experimental Results}

In order to track the average condition of the FSO link, we have implemented a simple SNR predictor based on a moving average over $N$ past measured SNR values,
\vspace{-0.35cm}
\begin{equation}
    \mathrm{SNR_{est}}(n+1) =   \frac{1}{N}\sum_{n-N+1}^{n}\mathrm{SNR_{meas}}(n) - \mathrm{SNR_{margin}},
    \label{eq:SNR}
\vspace{-0.35cm}
\end{equation}
where $\mathrm{SNR_{meas}}(n)$ is the measured SNR and \mbox{$\mathrm{SNR_{est}}(n+1)$} is the estimated SNR for the next time instant. 
In order to account for the variance of the FSO channel, a given SNR margin, $\mathrm{SNR_{margin}}$, must also be considered in the estimation. 
In a first phase of this work, we have analyzed the long-term average SNR behavior of the FSO link, from which we obtained an optimum number of taps of $N=3$, with an SNR margin of 2~dB.
Note that the optimized number of taps, $N$, captures the average time correlation of the channel, and therefore it depends on the sampling period between iterations. In this work, in the absence of real-time processing, each iteration, $n$, took an average of 25 seconds, already including the full processing of the adaptive modulation scheme.
Afterwards, we have started a continuous measurement with the high-capacity transmission system depicted in Fig.~\ref{fig:expSetup}, for a period of 3-hours. In each iteration, the oscilloscope obtains a batch of $2\times 10^{5}$ samples for 3 distinct modulation schemes using PCS-64QAM: i) fixed modulation at 400~Gbps, ii) fixed modulation at 500~Gbps and iii) adaptive modulation with the bit-rate adapted to the estimated SNR. 
The bit-rate adaptation is driven by the achievable information rate (AIR) for the estimated SNR considering a threshold NGMI, $\mathrm{NGMI_{th}}=0.9$, which is a typical value that guarantees post-FEC error-free transmission for most non-ideal SD-FEC algorithms \cite{NGMI-2018} with 20\% overhead, 
\vspace{-0.18cm}
\begin{equation}
    \mathrm{AIR}(n) = \Psi\left(\mathrm{SNR_{est}}(n),\mathrm{NGMI_{th}},M_\mathrm{PCS}\right),
    \vspace{-0.18cm}
\end{equation}
where $M_\mathrm{PCS}=64$ is the constellation size of the PCS template and $\Psi(\cdot)$ is a nonlinear function that has been previously computed via simulation and stored in a look-up table. 
The net bit-rate is then directly obtained from the AIR at each iteration as $R_b(n) = \mathrm{AIR}(n)R_s R_\mathrm{FEC} R_\mathrm{PIL}$, where $R_s=64$~Gbaud is the gross symbol-rate, $R_\mathrm{FEC}=5/6$ is the FEC rate and $R_\mathrm{PIL}=15/16$ is the DSP pilot-rate. Since both the 400~Gpbs and 500~Gbps configurations are fixed and do not depend on the SNR estimation, their captured waveforms are only recorded during the measurement period and processed afterwards. 

\newlength{\figH}
\setlength{\figH}{4.75cm}
\begin{figure*}[t!]
\centering
\subfloat{
\includegraphics[height = \figH]{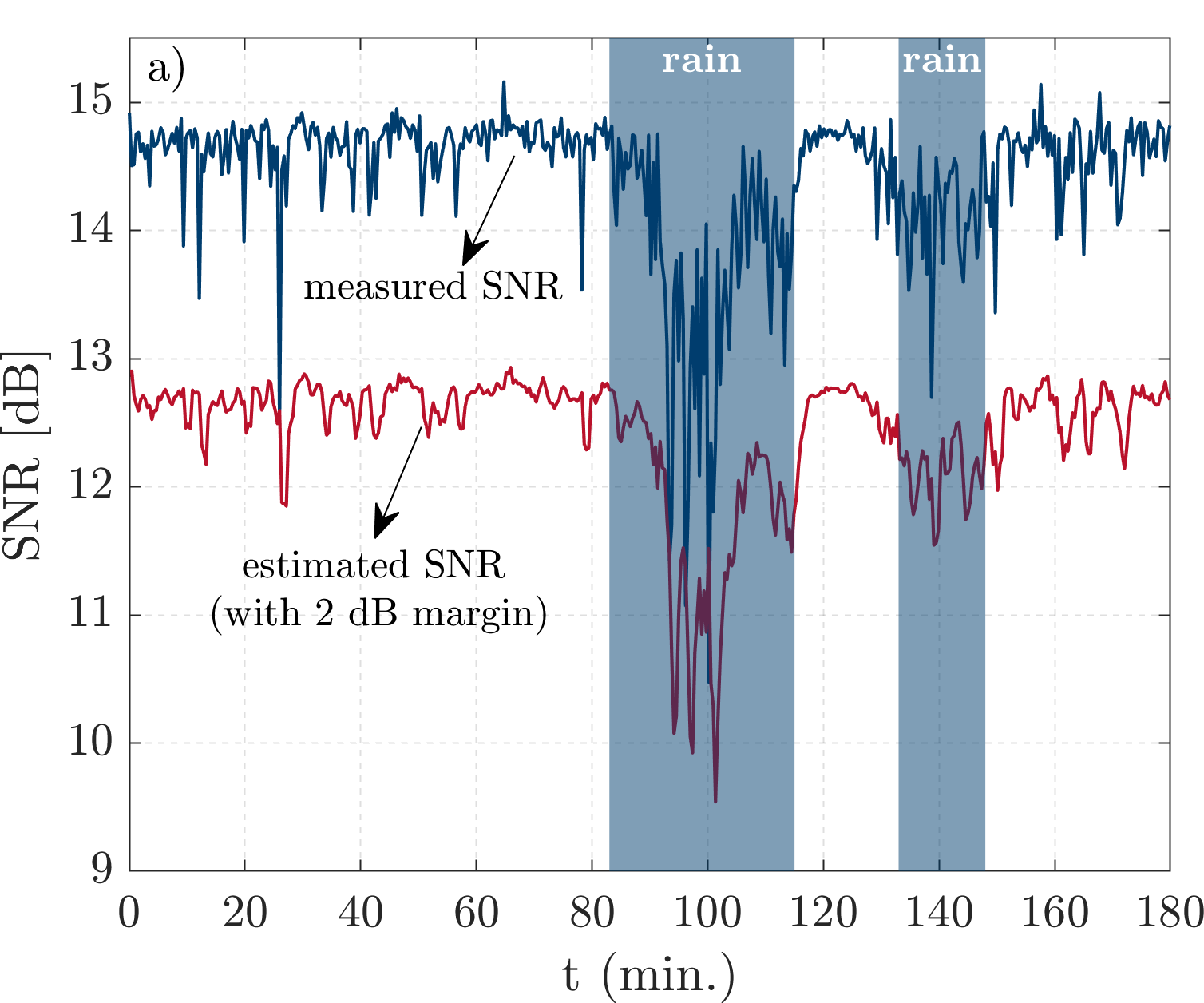}
\label{fig:SNR-vs-t}
}
\subfloat{
\includegraphics[height = \figH]{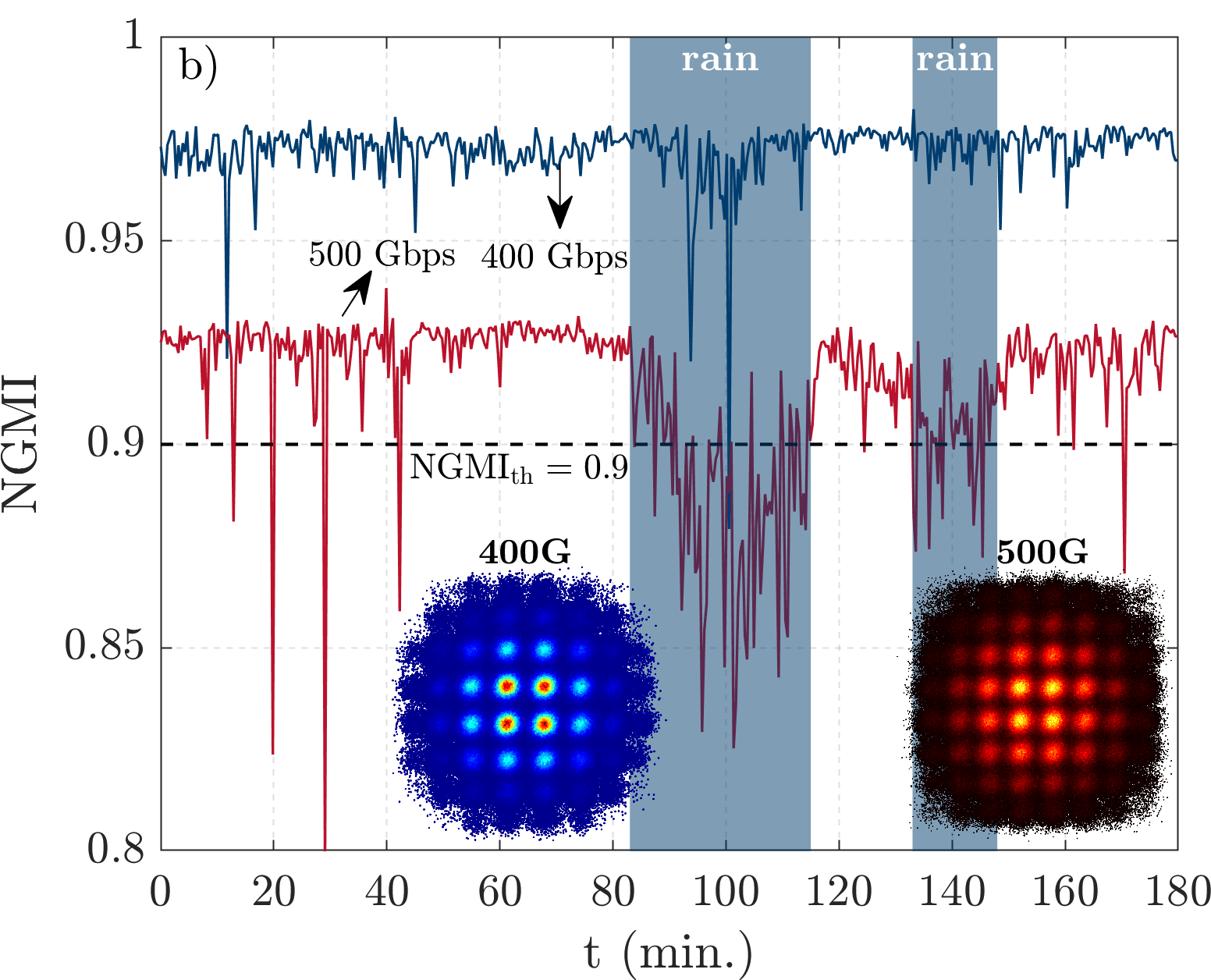}
\label{fig:NGMI-vs-t}
}
\subfloat{
\hspace{0.2cm}
\includegraphics[height = \figH]{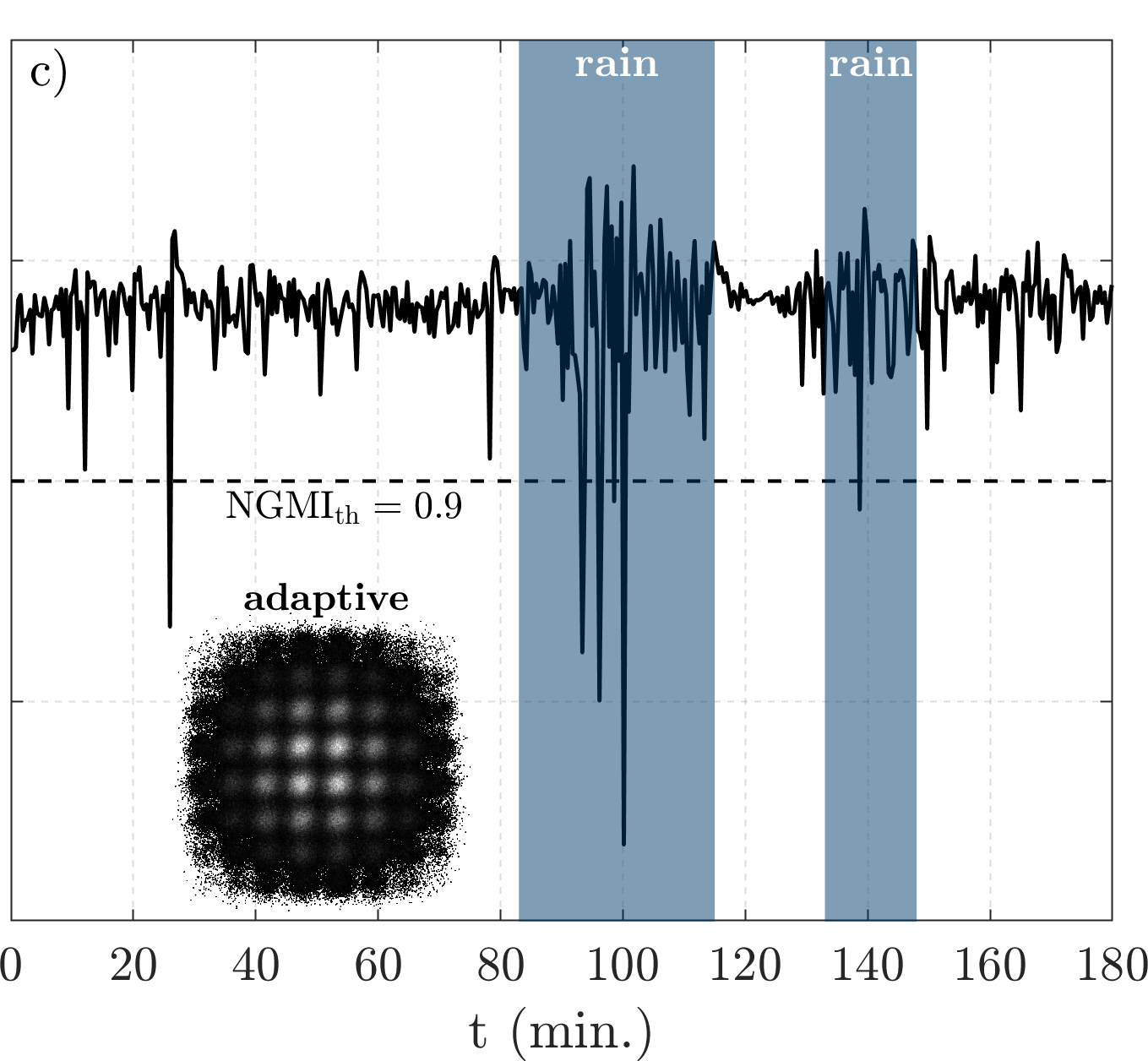}
\label{fig:NGMI-vs-t-adaptive}
}\\
\vspace{0.0cm}
\setlength{\figH}{4.3cm}
\subfloat{
\includegraphics[height = \figH]{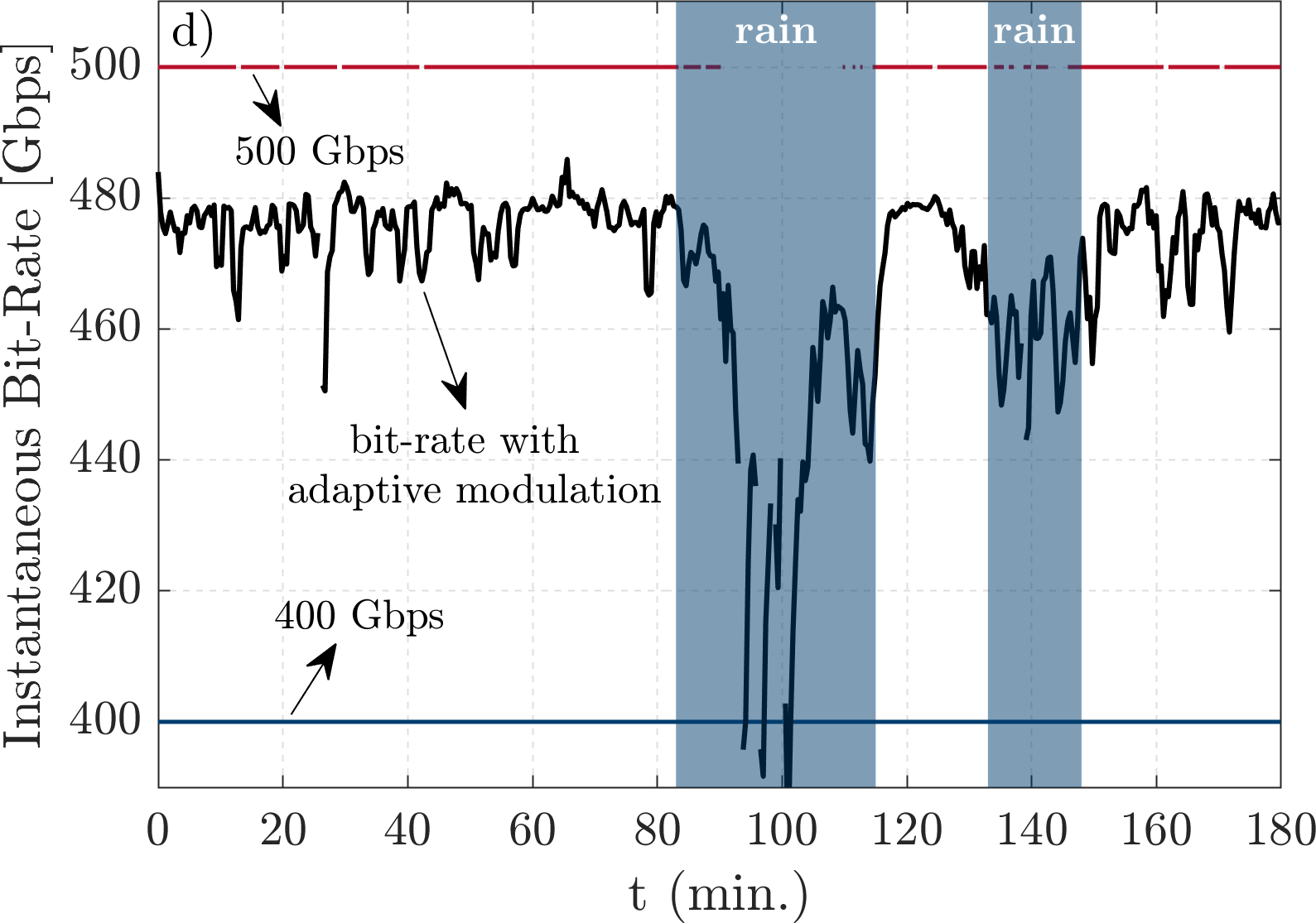}
\label{fig:Rb-vs-t}
}
\subfloat{
\includegraphics[height = \figH]{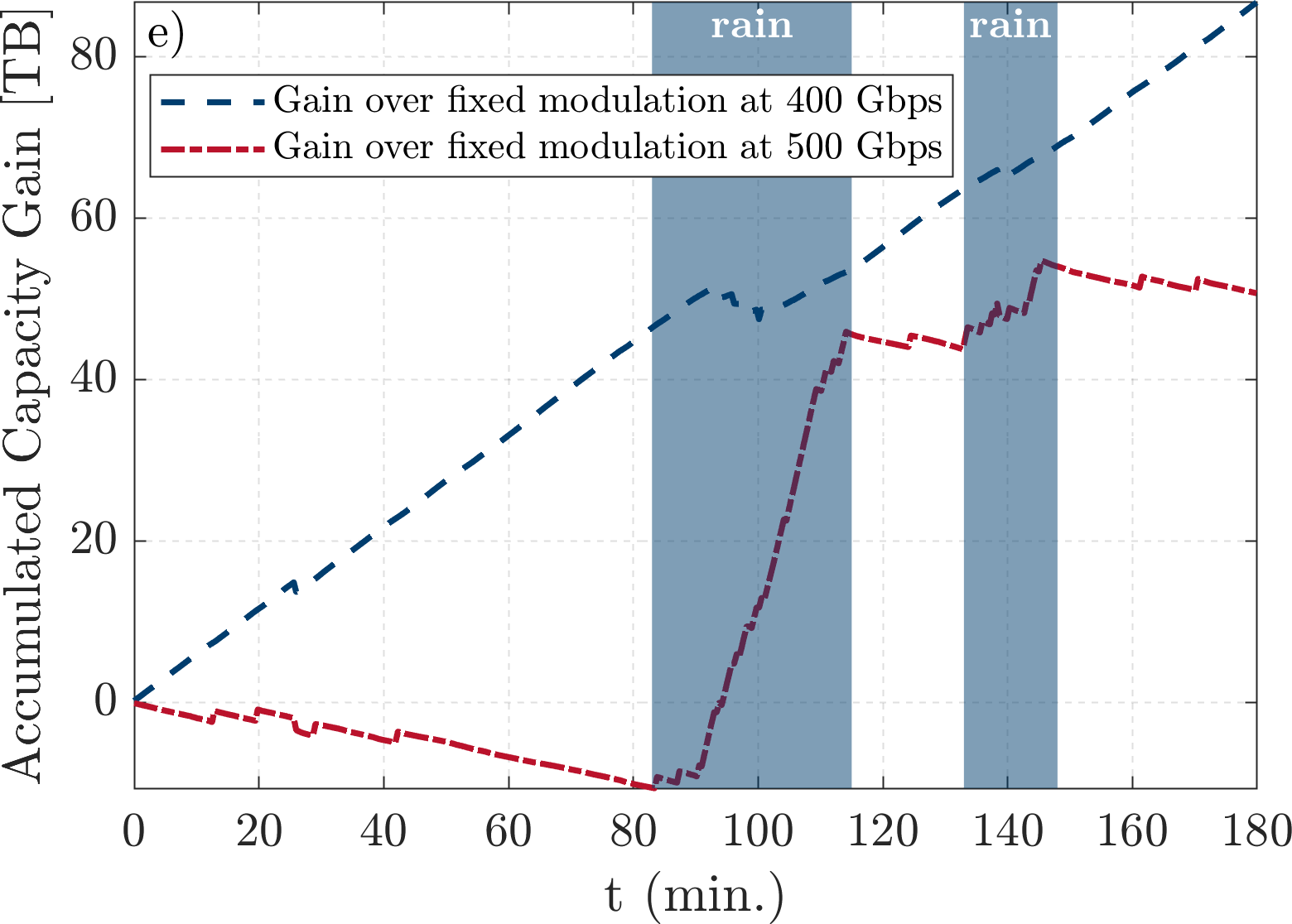}
\label{fig:C-vs-t}
}
\caption{(a) Measured SNR from the processed received signal and corresponding estimated SNR following the prediction rule \eqref{eq:SNR}; (b) obtained NGMI when the transmitted PCS-64QAM signal carries 400~Gbps and 500~Gbps; (c) obtained NGMI when adaptive PCS modulation is applied; (d) instantaneous transmitted bit-rate with fixed and adaptive modulation and (e) accumulated capacity gain (in Terabytes) obtained by time-adaptive modulation over fixed modulation.}
\label{fig:results}
\end{figure*}

The obtained results over 3 hours of acquisition time are depicted in Fig.~\ref{fig:results}, where the raining periods are highlighted in the blue shaded areas. The measured SNR and the respective estimated SNR (with 2~dB margin) are shown in Fig.~\ref{fig:SNR-vs-t}, where the impact of the rain is found to significantly reduce the average SNR of the channel. The red curve clearly shows that the simple SNR predictor of expression \eqref{eq:SNR} is able to accurately estimate the evolving condition of the FSO channel. 
In Fig.~\ref{fig:NGMI-vs-t}, we can see due to the more stringent requirements in terms of SNR, the fixed 500G solution is strongly affected during the raining periods, becoming out-of-service (i.e. below the NGMI threshold) for long periods of time. On the contrary, the 400G solution seems to be quite robust, even to the most adverse conditions, due to its inherently large SNR margin. However, this also means that for most the time the FSO channel is being under-utilized. 
Indeed, an improved utilization of the time-varying FSO channel can be achieved by the adaptive modulation scheme. As shown in Fig.~\ref{fig:NGMI-vs-t-adaptive}, the average NGMI is more effectively stabilized over the whole duration of the measurement. 
Nevertheless, the higher variance of the channel during the raining periods, which is likely to be due to the scattering of light when passing through raindrops, reflects into a higher variance of NGMI, still leading to a few violations of the NGMI threshold. 
The instantaneous bit-rates over time are shown in Fig.~\ref{fig:Rb-vs-t}, where the evolution of the adaptative bit-rate strategy with the varying weather condition can be clearly observed. 
While 400G provides a conservative but resilient solution, yielding an overall average bit-rate of 399.8~Gbps over the 3-hours measurement period, 500G fits to an optimistic estimation of the channel condition, enabling a very-high throughput during the best SNR periods, but failing catastrophically during the raining periods, which leads to an average of 426.6~Gbps, discarding the data in the periods in which the NGMI is below the 0.9 threshold. 
In comparison, the adaptive bit-rate strategy effectively compensates for the worst channel condition with a reduction of transmitted bit-rate, thereby relaxing the SNR requirements, which minimizes the losses of signal. Consequently, an improved average bit-rate of 464~Gbps is achieved over the 3-hours measurement period. 
The accumulated capacity gain of the adaptive modulation scheme relatively to the fixed 400G and 500G strategies is then summarized in Fig.~\ref{fig:C-vs-t}, in which we can observe a steady gain increase over the 400G signal, except for a short interlude during the raining period. Overall, the total estimated gain after the 3-hours measurement is $>$80~Terabytes. 
Regarding the 500G signal, we can observe a slight capacity loss of the adaptive modulation during the first 80 minutes, which is due to the large considered SNR margin of 2~dB. Nevertheless, this initial loss is largely compensated by the rapid gain during the raining periods, leading to a total gain of $>$50~Terabytes after 3 hours.

    \section{Conclusions}
\vspace{-0.3cm}
Using a simple and low-complexity algorithm for channel estimation together with time-adaptive probabilistic constellation shaping, we have demonstrated significant capacity gains of $>64$~Gbps and $>37$~Gbps over fixed PCS modulation at 400G and 500G, respectively. 
In addition, the obtained results provide further evidence of the robustness of free-space optics against rainy weather conditions.
    
    \section{Acknowledgements}

\vspace{-0.3cm}
\small
This work was partially supported by the European Regional Development Fund (FEDER), through the Regional Operational Programme of Centre (CENTRO 2020) of the Portugal 2020 framework, through the projects ORCIP (CENTRO-01-0145-FEDER-022141), 5GO (POCI-01-0247-FEDER-024539), SOCA (CENTRO-01-0145-FEDER-000010), and LandMark (PTDC/EEI-COM/31527/2017,  SAICT-45-2017-POCI-01-0145-FEDER-031527).


\end{document}